\documentclass[preprint,aos]{imsart}

\RequirePackage{amsthm,amsmath,amsfonts,amssymb}
\RequirePackage{mathtools}
\RequirePackage[authoryear]{natbib}
\RequirePackage[colorlinks,citecolor=blue,urlcolor=blue]{hyperref}
\RequirePackage{graphicx}

\RequirePackage{multirow}
\RequirePackage{tabulary}

\startlocaldefs

\theoremstyle{plain}
\newtheorem{proposition}{Proposition}

\newtheorem{theorem}{Theorem}
\newtheorem{lemma}{Lemma}

\theoremstyle{definition}
\newtheorem{remark}{Remark}
\newtheorem{assumption}{Assumption}

\newcommand{\CI}{{\cal I}}

\newcommand{\dee}{\mbox{d}}

\newcommand{\law}{{\cal L}}

\newcommand{\SE}{\mathbb{E}}
\newcommand{\SP}{\mathbb{P}}
\newcommand{\SR}{\mathbb{R}}

\newcommand{\trans}{^\top}

\DeclareMathOperator{\var}{Var}

\newcolumntype{C}[1]{>{\centering\arraybackslash}p{#1}}

\endlocaldefs

\begin{document}

\begin{frontmatter}

\title{Testing for conditional independence\\ in binary single-index models}
\runtitle{Testing conditional independence}

\begin{aug}
\author{\fnms{John H. J.}~\snm{Einmahl}},
\author{\fnms{Denis}~\snm{Kojevnikov}} 
\and
\author{\fnms{Bas J. M.}~\snm{Werker}}
\runauthor{J. H. J. Einmahl, D. Kojevnikov \and B. J. M. Werker}
\address{Department of Econometrics \& Operations Research, Tilburg University}
\end{aug}

\begin{abstract}
We wish to test whether a real-valued variable $Z$ has explanatory power, in addition to a multivariate variable $X$, for a binary variable $Y$.  Thus, we are interested in testing the hypothesis $\SP(Y=1\, | \, X,Z)=\SP(Y=1\, | \, X)$, based on $n$ i.i.d.\ copies of $(X,Y,Z)$. In order to avoid the curse of dimensionality, we follow the common approach of assuming that the dependence of both $Y$ and $Z$ on $X$ is through a single-index $X^\top\beta$ only. Splitting the sample on both $Y$-values, we construct a two-sample empirical process of transformed $Z$-variables, after splitting the $X$-space into parallel strips. Studying this two-sample empirical process is challenging: it does not converge weakly to a standard Brownian bridge, but after an appropriate normalization it does. We use this result to construct distribution-free tests.
\end{abstract}



\end{frontmatter}


\section{Introduction}\label{sec:Introduction}

Consider a binary variable $Y\in\{0,1\}$ and a multivariate explanatory variable $X\in\SR^d$. We also observe an additional real-valued variable $Z\in\SR$. We wish to test that $Z$ has explanatory power for $Y$ in addition to $X$, i.e.,
$\SP\left\{Y=1\vert X,Z\right\} = \SP\left\{Y=1\vert X\right\}$. To test this hypothesis, we assume that we are given an i.i.d.\ sample $(X_i,Y_i,Z_i)_{i=1}^n$. Our model setup is introduced more precisely in Section~\ref{sec:Model}.

Note that the hypothesis above is equivalent to the statement that, conditionally on $X$, the binary variable $Y$ and real-valued variable $Z$ are independently distributed. As a result, our paper contributes to a large literature on tests for conditional independence. Testing conditional independence is a statistically hard problem. \cite{shah_peters_2020} formalizes this. Their no-free-lunch Theorem~2 shows that, in case $(Y,X,Z)$ is only known to be absolutely continuously distributed, the power of any valid level $\alpha$ test is bounded by $\alpha$ as well. Thus, a valid test cannot have power against \emph{any} alternative. This result shows that practically useful conditional independence tests are always designed for more specific settings, effectively reducing the class of data generating processes for which validity of the test is required.

We address the no-free-lunch problem by considering only a binary variable $Y$ and by assuming a single-index structure: for some $\beta\in\SR^k$, the conditional distributions of $Y$ and $Z$ given $X$ depend on $X^\top\beta$ only. The assumed single-index structure allows us to circumvent the curse of dimensionality when estimating distributions conditional on $X$. The single-index assumption \citep{powell_1989} is often used for this reason in the context of testing for conditional independence, for instance by \cite{song_2009}.

Various approaches exist to testing conditional independence. For instance, one may rewrite the hypothesis to test in terms of multiplicativity of conditional characteristic functions \citep{su_white_2007} or conditional densities \citep{su_white_2008}. Another approach uses Rosenblatt transforms, see, e.g., \cite{cai_li_zhang_2022} and \cite{song_2009}. For instance, \cite{berrett_2020} uses concepts of permutation tests. Summarizing, the literature on conditional independence tests is vast. We refer to \cite{li_fan_2019} for a relatively recent comprehensive survey.

It is important to realize that, in view of the no-free-lunch theorem by \cite{shah_peters_2020}, all approaches, either explicitly or implicitly, make assumptions restricting applicability to specific applications only. An applied researcher thus always faces the choice between a multitude of possible tests. Our test is applicable in situations where $Y$ is binary and the single-index assumption can be expected to hold.

The test we propose in Section~\ref{sec:Test} is based on comparing the empirical conditional distribution functions of $Z$ given the single index $X^\top\beta$ for the subsamples with $Y=0$ and $Y=1$ separately. Importantly, this leads to distribution-free tests, i.e., tests with a constant size over the entire null hypothesis. Distribution freeness is not only computationally convenient, but generally also improves power of tests. Typically, conditional independence tests are not distribution free, notable exceptions being \cite{song_2009} and \cite{cai_li_zhang_2022}.

The remainder of this paper is organized as follows. Section~\ref{sec:Model} formally introduces our modeling setup. Section~\ref{sec:Test} contains the main weak convergence result that forms the basis of our test. Section~\ref{sec:Simulations} shows, by means of a small simulation study, that our asymptotic theory provides a good approximation for finite samples and investigates the power of our test. All proofs are gathered in Section~\ref{sec:Proofs}.

\section{Model setup}\label{sec:Model}

\noindent We wish to test whether a real-valued variable $Z$ has explanatory power, in addition to a $d$-variate variable $X$, for a binary variable $Y\in\{0,1\}$. Thus, we are interested in the hypothesis
\begin{equation}\label{eq:NullNonSingleIndex}
\SP\left\{Y=1\vert X,Z\right\} = \SP\left\{Y=1\vert X\right\},
\end{equation}
where equality between the conditional probabilities is to be understood in the almost sure sense. To test~(\ref{eq:NullNonSingleIndex}), we are given $n$ observations $(X_i,Y_i,Z_i)_{i=1}^n$ that are i.i.d.\ copies of $(X,Y,Z)$.

In order to avoid the curse of dimensionality, we follow the common approach of assuming that the dependence of both $Y$ and $Z$ on $X$ is through a single-index vector $\beta\in\SR^d$ only. That is, we assume that there exists a $\beta_0\in\SR^d$, with $||\beta_0||=1$, such that
\begin{eqnarray}
\label{eq:SingleIndexY}
\SP\left\{Y=1\vert X\right\} &=& \SP\left\{Y=1\vert X\trans\beta_0\right\},\\
\label{eq:SingleIndexZ}
\SP\left\{Z\leq z\vert X\right\} &=& \SP\left\{Z\leq z\vert X\trans\beta_0\right\}\mbox{ for all }z\in\SR.
\end{eqnarray}
Without these relations, the proofs would require contradictory conditions on the growth rate of the  number of ``cells'' $m$, defined in~(\ref{eq:AkDef}) below, because in general the diameter of the cells would roughly decrease as slowly as $1/m^{1/d}$.

Given the single-index structure~(\ref{eq:SingleIndexY})--(\ref{eq:SingleIndexZ}), we  specify the null hypothesis that $Z$ has no additional explanatory power for $Y$ as
\begin{equation}\label{eq:NullHypothesis}
H_0:~~~\SP\left\{Y=1\vert X,~Z\right\} = \SP\left\{Y=1\vert X\trans\beta_0\right\}.
\end{equation}
Note that~(\ref{eq:NullHypothesis}) in itself already implies the single-index structure~(\ref{eq:SingleIndexY}) for $Y$. Moreover, as $Y$ is binary, (\ref{eq:SingleIndexY}) and~(\ref{eq:NullHypothesis}) imply the equivalent statements for $Y=0$.
\begin{remark}
Note that omitting a variable in a binary choice model does not necessarily make the model misspecified. For instance, suppose that $\Phi$ denotes the standard normal cumulative distribution function and we have
\begin{equation}\nonumber
\SP\left\{Y=1\vert X,Z\right\}
 =
\Phi\left(\alpha+\beta X+\gamma Z\right),
\end{equation}
for some $\alpha$, $\beta$, and $\gamma\neq0$. If $X$ and $Z$ are jointly normally distributed, this also implies
\begin{equation}\nonumber
\SP\left\{Y=1\vert X\right\}
 =
\Phi\left(\delta_0+\delta_1 X\right),
\end{equation}
for some $\delta_0$ and $\delta_1$. Thus, in this particular case, omitting the variable $Z$ from the analysis does not induce misspecification. However, correctly taking $Z$ into account does lead to more discrimination between $Y=0$ and $Y=1$.
\end{remark}

The following lemma provides two implications of~(\ref{eq:SingleIndexZ}) and~(\ref{eq:NullHypothesis}) that allow us to test our null hypothesis of interest within the single-index setting, thereby avoiding the curse of dimensionality.

\begin{lemma}
	Assume that~(\ref{eq:SingleIndexZ})--(\ref{eq:NullHypothesis}) hold. Then, we have~(\ref{eq:NullNonSingleIndex}) or, equivalently,
	\begin{equation}\label{eq:NullHypothesisRewritten1}
		\law\left(Z\vert X,Y\right) = \law\left(Z\vert X\right).
	\end{equation}
	Moreover, we then have
\begin{equation}\label{eq:NullHypothesisRewritten2}
	\law\left(Z\vert X,Y\right) =
    \law\left(Z\vert X\trans\beta_0\right) =
    \law\left(Z\vert X\trans\beta_0,Y\right).
	\end{equation}
\end{lemma}
\noindent{\bf Proof.}
Note that~(\ref{eq:NullHypothesis}) implies~(\ref{eq:SingleIndexY}) and, thus, (\ref{eq:NullNonSingleIndex}) and (\ref{eq:NullHypothesisRewritten1}).
To show~(\ref{eq:NullHypothesisRewritten2}), note, for $z\in\SR$,
\begin{equation*}
	\qquad\qquad\qquad\SP\left\{Z\leq z\vert X,~Y\right\}
\stackrel{(\ref{eq:NullHypothesisRewritten1})}{=}
	\SP\left\{Z\leq z\vert X\right\}
	\stackrel{(\ref{eq:SingleIndexZ})}{=}
	\SP\left\{Z\leq z\vert X\trans\beta_0\right\}.\qquad\qquad\qquad\Box
\end{equation*}
Note that~(\ref{eq:NullHypothesisRewritten1}) states that, conditionally on $X$, the variables $Y$ and $Z$ are independent. Relation~(\ref{eq:NullHypothesisRewritten2}) formalizes the single-index structure that is needed to avoid the curse of dimensionality.

\section{Testing for irrelevance of \texorpdfstring{$Z$}{Z}}\label{sec:Test}

\noindent We use the omitted variable hypothesis rewritten in~(\ref{eq:NullHypothesisRewritten1})-(\ref{eq:NullHypothesisRewritten2}) to construct our test statistic. More precisely, given the single-index structure, we will test equality of the conditional distribution of $Z$ given $X\trans\beta_0$ for the two subsamples defined by $Y=0$ and $Y=1$.
We will use:
\begin{assumption}\label{a1}
$X$ has bounded convex support $S$. The distribution of $X$ admits a density with respect to Lebesgue measure that is bounded away from zero on $S$.
\end{assumption}

Our  sample  consists of $n$ independent copies of  $(X,Y, Z)$: $(X_1,Y_1, Z_1), \dots, (X_n,Y_n, Z_n)$. We create a partition of the observations into $m$ subsamples on the basis of their value of $X$. More precisely, given $\beta\in\SR^d$ with $||\beta||=1$ and $m+1=m(n)+1$ numbers $a_0<a_1<\ldots<a_m$ 
such that $\mathbb{P} (a_0 <X^\top\beta \leq a_m)=1$, we define, for $k=1,\ldots,m$,
\begin{equation}\label{eq:AkDef}
	A_k = A_k(\beta) = \left\{x\in S:~a_{k-1}<x\trans\beta \leq a_k\right\}.
\end{equation}
Write $$ F_{j,k}(z) =\mathbb{P}(Z\leq z\, | \, X\in A_k, Y=j), \quad z\in \mathbb{R}.$$ 
We require:
\begin{assumption}\label{a2}
For all $k=1, \dots, m$ and $j=0,1$, the $F_{j,k}$ are continuous and $m\mathbb{P}( X \in A_k, Y=j)$ stays bounded away from zero, as $m\to \infty$.
\end{assumption}

Given the partition $\left(A_k\right)_{k=1}^m$, we define the index sets
\begin{equation}\label{eq:IjkDef}
	\CI_{j,k} =	\left\{i=1,\ldots,n:~Y_i=j,~X_i\in A_k\right\}.
\end{equation}
Further, we denote by $n_{j,k}=|\CI_{j,k}|$ the number of observations in $\CI_{j,k}$, i.e., $n_{j,k}$ is the number of observations for which $Y$ takes the value $j$ and $X\in A_k$. 
The total number of observations for which $Y$ takes the value $j$ is given by
\begin{equation}\label{eq:njDef}
	n_j = \sum_{k=1}^m n_{j,k}.
\end{equation}
Obviously, $n=n_0+n_1$.

To test equality of the distributions of $Z$ conditionally on $Y=0$ and $Y=1$, given $X\in A_k$, we introduce the empirical distribution functions $\widehat{F}_{j,k}$, for $j=0,1$ and $k=1,\ldots,m$, defined by, for $z\in\SR$,
\begin{equation}\label{eq:hatFjkDef}
	\widehat{F}_{j,k}\left(z\right) = \frac{1}{n_{j,k}}\sum_{i\in\CI_{j,k}}I\left\{Z_i< z\right\},
\end{equation}
and the combined empirical distribution functions $$ \widehat F_k(z)=\frac{n_{0,k}}{n_{0,k}+n_{1,k}}\widehat F_{0,k}(z)+\frac{n_{1,k}}{n_{0,k}+n_{1,k}}\widehat F_{1,k}(z). $$

Key ingredients in our test statistic are the subsample empirical distribution functions $\widehat{\Gamma}_j$, $j=0,1$, defined by, for $u\in[0,1]$,
\begin{equation}\label{eq:hatGammajDef}
	\widehat{\Gamma}_j\left(u\right) = 
	\frac{1}{n_j}\sum_{k=1}^m\sum_{i\in\CI_{j,k}}I\left\{
	\widehat{F}_{k}\left(Z_i\right)< u\right\}.
\end{equation}

We now  consider the situation where the $A_k$ are based on an estimator $\widehat \beta$ of $\beta_0$ (instead of on $\beta$) and hence random.
\begin{assumption}\label{a3}
Let $\check \beta_n$ be an estimator of $\beta_0$ such that $\sqrt{n}\check \beta_n\in \mathbb{Z}^d$ and $\check \beta_n-\beta_0=O_\mathbb{P}(1/\sqrt{n})$.
\end{assumption}
Given the single-index structure in~(\ref{eq:SingleIndexY})--(\ref{eq:SingleIndexZ}), root-$n$ consistent estimators of $\beta_0$ are easily constructed, see, e.g., \cite{powell_1989}. Moreover, if we have an estimator $\widetilde \beta_n$ with $\widetilde \beta_n-\beta_0=O_\mathbb{P}(1/\sqrt{n})$, then we can find  $\check \beta_n$  by rounding  all components on a $1/\sqrt{n}$ scale. Define $\widehat \beta=\check \beta_n/||\check \beta_n||$. Under the null hypothesis in (\ref{eq:NullHypothesis}), we expect $\widehat{\Gamma}_0$ and $\widehat{\Gamma}_1$ to be close (assuming that the partition $\left(A_k\right)_{k=1}^m$ is sufficiently fine; this will be made precise later). Therefore, define the process $\gamma_n$ by, for $u\in[0,1]$,
\begin{equation}\label{gam}
	\gamma_n(u) =
	\frac{\sqrt{\frac{n_0n_1}{n}}\,(
	\widehat{\Gamma}_0(u)-\widehat{\Gamma}_1(u))}{
	\sqrt{1-\frac{n_0n_1}{n}\sum_{k=1}^m
		\frac{(n_{0,k}/n_0-n_{1,k}/n_1)^2}{n_{0,k}+n_{1,k}}}}.
\end{equation}
We show that the limiting distribution of $\gamma_n$ under the null hypothesis is a standard Brownian bridge.
Note that the expression in the numerator mimics the two-sample empirical process, but the novel standardization through the denominator is needed to deal with the empirical distribution functions $\widehat{F}_{k}$ inside the indicators in the definitions of $\widehat \Gamma_0$ and $\widehat \Gamma_1$ in (\ref{eq:hatGammajDef}).
Our test statistic is then, finally, given by 
the Cram\a'{e}r-von Mises-type statistic
\begin{equation}\label{eq:CvM}
	T_n = \int_0^1\gamma_n^2(u)\dee u.
\end{equation}
Let `$\rightsquigarrow$' denote weak convergence and write $\ell^{\infty}([0,1])$ for  the space of bounded real functions on $[0,1]$.
Let $B$ be a standard Brownian bridge. The following theorem is our main result; its proof can be found in Section~\ref{sec:Proofs}.
\begin{theorem}\label{t1}
Let Assumptions~\ref{a1}--\ref{a3} hold and assume that
the (differentiability) conditions of Lemma~\ref{lem3} below are satisfied.
If $$(m^3\log^3 n) /n\to 0\,\,\, \mbox{ and } \,\,\, m^4/n\to \infty, \quad \mbox{as } n\to \infty,$$ then, 
under the null hypothesis in~(\ref{eq:NullHypothesis}), we have 
  \begin{equation*}
			\gamma_n\rightsquigarrow B \quad \text{in }  \ell^{\infty}([0,1]), \quad \mbox{as }  n \to \infty,
		\end{equation*}
        and, consequently,
$$T_n\stackrel{d}{\to}\int_0^1B^2(u)\dee u.   
\quad 
$$        
\end{theorem}
Observe that the limit $B$ of the ``test process'' $\gamma_n$ is not only distribution-free but also a well-known Gaussian process. As a consequence, the usual test statistics based on $\gamma_n$ have well-studied distributions. Based on favorable simulation results, we use the Cram\a'{e}r-von Mises-type statistic, but one could equally well use the Kolmogorov-Smirnov-type statistic or any other continuous functional of $\gamma_n$.

\section{Simulation analysis}\label{sec:Simulations}

In this section, we investigate the finite-sample performance of the proposed Cram\a'{e}r-von Mises-type test statistic~(\ref{eq:CvM}) through Monte Carlo simulations. We examine the empirical size under the null hypothesis and the power under various alternatives, and assess how well the asymptotic distribution approximates the finite-sample distribution.

\subsection*{Data generating process}
We consider the following flexible data generating process. The covariate vector $X$ follows a uniform distribution on $[(-1, 1)^d]$ with $d \in \{3, 5\}$. The variable $Z$ is generated as 
\[
    Z = \exp(X^\top\beta) + (1 + |X^\top\beta|)^\sigma U,
\]
where $U \sim \mathcal{N}(0,1)$ is independent of $X$. The binary response is defined as $Y = 1\{X^\top\beta + \theta U > V\}$, where $V \sim \mathcal{N}(0,1)$ is independent of $(X, U)$. We take $\beta = \mathbf{1}_d$ and set $\beta_0=\beta/\sqrt{d}$.

Under this design, the null hypothesis $H_0: \SP\{Y = 1 \mid X, Z\} = \SP\{Y = 1 \mid X^\top\beta_0\}$ holds if and only if $\theta = 0$. When $\theta \neq 0$, the variable $Z$ has additional explanatory power for $Y$ beyond $X^\top\beta$, and the null hypothesis is violated. The parameter $\sigma \in \{0, 1/4, 1/2, 3/4, 1\}$ controls the heteroskedasticity of $Z$ given $X$, with larger values of $\sigma$ inducing stronger heteroskedasticity.

We set the sample size to $n = 1{,}000$ and consider $m \in \{10, 15, 20\}$ cells for partitioning the $X$-space. The cells are constructed to have about equal empirical probability mass under the estimated single-index structure. All results are based on $5{,}000$ Monte Carlo replications, and we use a nominal significance level of $5\%$.

We implement the test using two approaches for constructing the partition. First, as a benchmark, we use the oracle approach where the true index $X^\top\beta_0$ is employed to construct the cells $A_k$, $k=1,\ldots,m$. Second, following \citet{powell_1989}, we estimate the direction $\beta_0$ via the density-weighted average derivative
\[
    \delta=\SE[f_X(X)g'(X^\top\beta_0)]\beta_0,
\]
where $g(x) = \SP\{Y = 1 \mid X^\top\beta_0 = x\}$ and $f_X$ is the density of $X$. We denote the estimator by $\widehat{\delta}$, then normalize $\widehat{\beta} = \widehat{\delta}/\|\widehat{\delta}\|$ and use $X^\top\widehat{\beta}$ to construct the cells. This second approach represents the practical implementation of our test.

\subsection*{Size analysis}
Tables \ref{tab:size_d3} and \ref{tab:size_d5} report the empirical rejection rates under the null hypothesis ($\theta = 0$) for dimensions $d = 3$ and $d = 5$, respectively. For both implementations (oracle and average derivative), the empirical sizes are close to the nominal $5\%$ level across all values of $\sigma$ and $m\in\{15,20\}$.

Using the average derivative estimator instead of the oracle direction introduces minimal size distortion when $d=3$, suggesting that the estimation uncertainty is adequately handled by the asymptotic theory. For $d=5$, the test based on the average derivative estimator shows slightly more size distortion, especially at smaller values of $m$, which reflects the increased difficulty of estimating the direction in higher dimensions. However, at $m=20$, the test exhibits good size control, with empirical rejection rates consistently near the nominal level. 

\begin{table}[t]
\centering
\renewcommand{\arraystretch}{0.9}
\renewcommand{\baselinestretch}{1}
\caption{Empirical size of Cram\a'{e}r-von Mises test for $d = 3$ \\ ($n = 1{,}000$, nominal level $5\%$)}
\label{tab:size_d3}
\begin{tabular}{c|ccc|ccc}
\hline
\hline
& \multicolumn{3}{c|}{Oracle} & \multicolumn{3}{c}{Average Derivative} \\
$\sigma$ & $m=10$ & $m=15$ & $m=20$ & $m=10$ & $m=15$ & $m=20$ \\
\hline
0.00 & 0.067 & 0.052 & 0.052 & 0.066 & 0.056 & 0.051 \\
0.25 & 0.059 & 0.050 & 0.049 & 0.063 & 0.050 & 0.047 \\
0.50 & 0.056 & 0.049 & 0.045 & 0.064 & 0.058 & 0.053 \\
0.75 & 0.058 & 0.051 & 0.050 & 0.067 & 0.055 & 0.052 \\
1.00 & 0.054 & 0.052 & 0.054 & 0.056 & 0.050 & 0.048 \\
\hline
\hline
\end{tabular}
\end{table}
\begin{table}[t]
\centering
\renewcommand{\arraystretch}{0.9}
\renewcommand{\baselinestretch}{1}
\caption{Empirical size of Cram\a'{e}r-von Mises test for $d = 5$ \\ ($n = 1{,}000$, nominal level $5\%$)}
\label{tab:size_d5}
\begin{tabular}{c|ccc|ccc}
\hline
\hline
& \multicolumn{3}{c|}{Oracle} & \multicolumn{3}{c}{Average Derivative} \\
$\sigma$ & $m=10$ & $m=15$ & $m=20$ & $m=10$ & $m=15$ & $m=20$ \\
\hline
0.00 & 0.069 & 0.051 & 0.045 & 0.086 & 0.063 & 0.058 \\
0.25 & 0.068 & 0.052 & 0.053 & 0.075 & 0.056 & 0.052 \\
0.50 & 0.065 & 0.057 & 0.054 & 0.081 & 0.067 & 0.061 \\
0.75 & 0.062 & 0.056 & 0.054 & 0.071 & 0.058 & 0.054 \\
1.00 & 0.062 & 0.054 & 0.052 & 0.071 & 0.063 & 0.057 \\
\hline
\hline
\end{tabular}
\end{table}

\subsection*{Assessment of asymptotic approximation}
\begin{figure}[t]
    \centering
    \renewcommand{\baselinestretch}{1}
    \includegraphics[width=\textwidth]{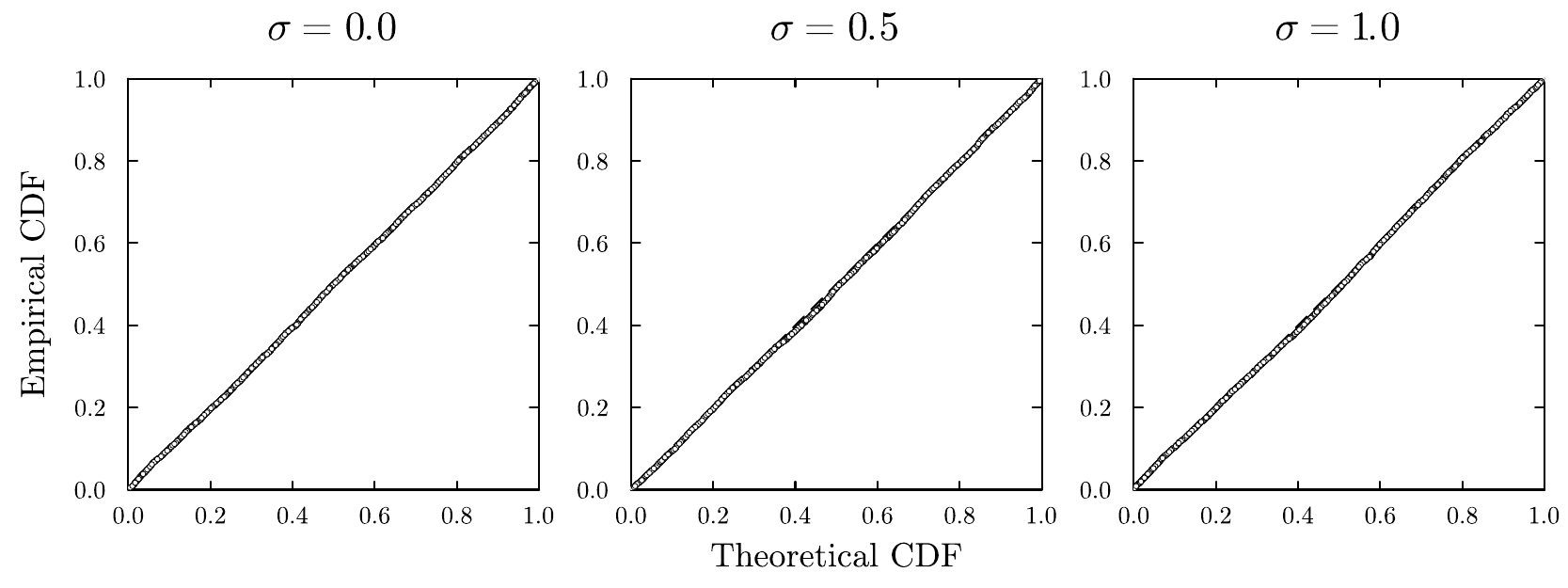}
    \caption{Probability-probability plots for Cram\a'{e}r-von Mises statistic for $d = 3$ using average derivative estimator ($n = 1{,}000$, $\theta = 0$, $m = 20$).}
    \label{fig:ppplots_d3}
\end{figure}

\begin{figure}[t]
    \centering
    \renewcommand{\baselinestretch}{1}
    \includegraphics[width=\textwidth]{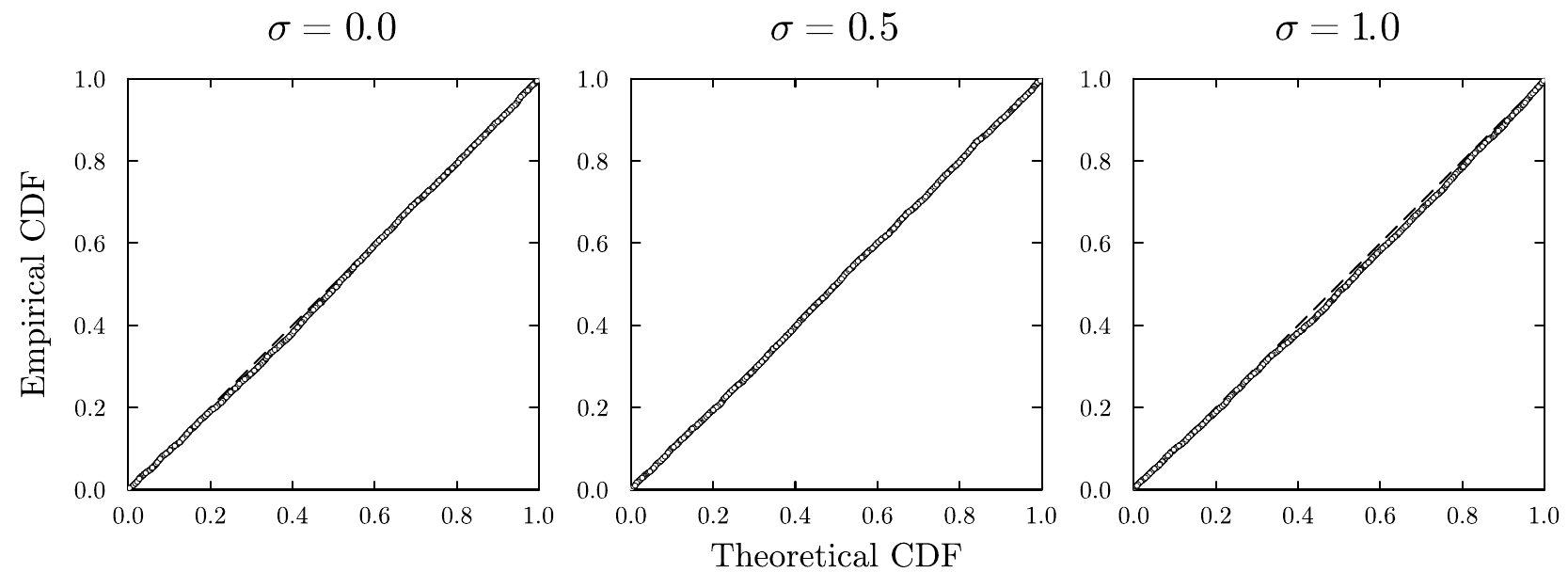}
    \caption{Probability-probability plots for Cram\a'{e}r-von Mises statistic for $d = 5$ using average derivative estimator ($n = 1{,}000$, $\theta = 0$, $m = 20$).}
    \label{fig:ppplots_d5}
\end{figure}

To evaluate how well the asymptotic distribution approximates the finite-sample distribution of the test statistic, we examine probability-probability plots, comparing the empirical cumulative distribution function of the Cram\a'{e}r-von Mises-type statistic with its asymptotic distribution under the null hypothesis. Figures \ref{fig:ppplots_d3} and \ref{fig:ppplots_d5} present probability-probability plots for $d = 3$ and $d=5$, respectively, with $m = 20$ and $\theta = 0$ (null hypothesis). Each figure displays results for the average derivative estimator, across three levels of heteroskedasticity: $\sigma \in \{0, 0.5, 1.0\}$.

We see that for both $d=3$ and $d=5$, the Cram\a'{e}r-von Mises-type statistic shows very good agreement with the asymptotic distribution across all values of $\sigma$. This  demonstrates that the
single-index structure effectively mitigates the curse of dimensionality that would severely
impact fully nonparametric approaches.

\subsection*{Power analysis}
Tables \ref{tab:power_d3} and \ref{tab:power_d5} display the empirical power of the Cram\a'{e}r-von Mises test using the average derivative estimator for $d=3$ and $d=5$, respectively. The tables show power across different numbers of cells $m \in \{10, 15, 20\}$, levels of heteroskedasticity $\sigma \in \{0.00, 0.25, 0.50, 0.75, 1.00\}$, and several values of $\theta \in \{-0.25, -0.10, 0.00, 0.10, 0.25\}$. The column corresponding to $\theta = 0.00$ reports the empirical size under the null hypothesis, while the remaining columns show power under various alternatives.

\begin{table}[t]
\centering
\renewcommand{\arraystretch}{0.9}
\renewcommand{\baselinestretch}{1}
\caption{Empirical power of Cram\a'{e}r-von Mises test for $d = 3$ using average derivative estimator \\ ($n=1{,}000$, nominal level $5\%$)}
\label{tab:power_d3}
\begin{tabular}{cc|C{4.5em}C{4.5em}C{4.5em}C{4.5em}C{4.5em}}
\hline
\hline
$m$ & $\sigma$ & $\theta=-0.25$ & $\theta=-0.10$ & $\theta=0.00$ & $\theta=0.10$ & $\theta=0.25$ \\
\hline
\multirow{5}{*}{10}
& 0.00 & 0.994 & 0.353 & 0.066 & 0.669 & 0.999 \\
& 0.25 & 0.994 & 0.348 & 0.063 & 0.671 & 1.000 \\
& 0.50 & 0.997 & 0.365 & 0.064 & 0.636 & 0.999 \\
& 0.75 & 0.997 & 0.373 & 0.067 & 0.633 & 1.000 \\
& 1.00 & 0.997 & 0.383 & 0.056 & 0.638 & 0.999 \\
\hline
\multirow{5}{*}{15}
& 0.00 & 0.998 & 0.428 & 0.056 & 0.595 & 1.000 \\
& 0.25 & 0.997 & 0.423 & 0.050 & 0.606 & 0.999 \\
& 0.50 & 0.997 & 0.440 & 0.058 & 0.579 & 0.999 \\
& 0.75 & 0.999 & 0.433 & 0.055 & 0.583 & 1.000 \\
& 1.00 & 0.998 & 0.432 & 0.050 & 0.589 & 0.999 \\
\hline
\multirow{5}{*}{20}
& 0.00 & 0.998 & 0.455 & 0.051 & 0.577 & 0.999 \\
& 0.25 & 0.998 & 0.449 & 0.047 & 0.582 & 0.999 \\
& 0.50 & 0.998 & 0.460 & 0.053 & 0.555 & 0.999 \\
& 0.75 & 0.998 & 0.457 & 0.052 & 0.556 & 1.000 \\
& 1.00 & 0.998 & 0.452 & 0.048 & 0.579 & 0.999 \\
\hline
\hline
\end{tabular}
\end{table}

\begin{table}[t]
\centering
\renewcommand{\arraystretch}{0.9}
\renewcommand{\baselinestretch}{1}
\caption{Empirical power of Cram\a'{e}r-von Mises test for $d = 5$ using average derivative estimator \\ ($n=1{,}000$, nominal level $5\%$)}
\label{tab:power_d5}
\begin{tabular}{cc|C{4.5em}C{4.5em}C{4.5em}C{4.5em}C{4.5em}}
\hline
\hline
$m$ & $\sigma$ & $\theta=-0.25$ & $\theta=-0.10$ & $\theta=0.00$ & $\theta=0.10$ & $\theta=0.25$ \\
\hline
\multirow{5}{*}{10}
& 0.00 & 0.975 & 0.249 & 0.086 & 0.650 & 0.999 \\
& 0.25 & 0.981 & 0.261 & 0.075 & 0.650 & 1.000 \\
& 0.50 & 0.985 & 0.272 & 0.081 & 0.646 & 1.000 \\
& 0.75 & 0.985 & 0.286 & 0.071 & 0.652 & 0.999 \\
& 1.00 & 0.983 & 0.277 & 0.071 & 0.647 & 0.999 \\
\hline
\multirow{5}{*}{15}
& 0.00 & 0.986 & 0.329 & 0.063 & 0.577 & 0.998 \\
& 0.25 & 0.990 & 0.338 & 0.056 & 0.576 & 0.999 \\
& 0.50 & 0.990 & 0.337 & 0.067 & 0.580 & 0.999 \\
& 0.75 & 0.991 & 0.350 & 0.058 & 0.596 & 0.999 \\
& 1.00 & 0.989 & 0.333 & 0.063 & 0.590 & 0.998 \\
\hline
\multirow{5}{*}{20}
& 0.00 & 0.990 & 0.357 & 0.058 & 0.539 & 0.998 \\
& 0.25 & 0.992 & 0.365 & 0.052 & 0.548 & 0.999 \\
& 0.50 & 0.992 & 0.364 & 0.061 & 0.555 & 1.000 \\
& 0.75 & 0.992 & 0.370 & 0.054 & 0.569 & 0.998 \\
& 1.00 & 0.990 & 0.347 & 0.057 & 0.569 & 0.998 \\
\hline
\hline
\end{tabular}
\end{table}

The test exhibits high power against substantial departures from the null hypothesis. For alternatives with $|\theta| = 0.25$, power exceeds $99\%$ in nearly all scenarios for both $d=3$ and $d=5$, regardless of the level of heteroskedasticity or the number of cells. This shows that the test reliably detects strong violations of conditional independence, even in dimension  5. For closer alternatives with $|\theta| = 0.10$, power ranges from approximately $25\%$ to $67\%$ depending on the configuration. The effects of  $m$ and $\sigma$ on  the power are generally modest. The asymmetry in power between positive and negative values of $\theta$ reflects the specific structure of the data generating process rather than a  property of the test.

\newpage

\section{Proofs}\label{sec:Proofs}

For the proof of the main theorem, we need some propositions and lemmas, which we present first. Throughout, we assume that the null hypothesis (\ref{eq:NullHypothesis}) holds.

Given $\beta \in \mathbb{R}^d$ with $||\beta||=1$, let $A_k,$ $ k=1, \ldots, m,$ be as in (\ref{eq:AkDef}). 
Recall $ F_{j,k}(z) =\mathbb{P}(Z\leq z\, | \, X\in A_k, Y=j)$ and $\widehat F_{j,k}(z)=\frac{1}{n_{j,k}}\sum_{i\in \mathcal{I}_{j,k}}  I\{Z_i<z\}.$ The following result follows from the Proposition in
\cite{beirlant1996maximal}.

\begin{proposition}\label{prop1}
Let Assumption~\ref{a2} hold. If $m\to \infty$ and $(m \log^2 n)/n\to 0$, as $n\to \infty$, then there exists a triangular scheme of  rowwise independent standard Brownian bridges $B_{j,k,n} , j\in \{0,1\},  k\in \{1, \ldots, m\}$, such that, as $n\to\infty$,
$$\max_{j\in \{0,1\}}\max_{k\in \{1, \ldots, m\}}\sup_{z\in \mathbb{R}}|\sqrt{n_{j,k}}(\widehat F_{j,k}(z)-F_{j,k}(z))-B_{j,k,n}(F_{j,k}(z))|=O_\mathbb{P} \left( \sqrt{\frac{m}{n}} \log n\right) .
$$\end{proposition}

Consider a process $\delta_n(\beta)=\delta_n(\cdot, \beta)$, on $[0,1]$, depending on $\beta = (\beta_1, \ldots, \beta_d)^\top$ that defines the partition into $A_1, \ldots, A_m$.    Write $\beta_{0,n}=(\lfloor\sqrt{n}\beta_1\rfloor/ \sqrt{n}, \ldots, \lfloor\sqrt{n}\beta_d\rfloor/ \sqrt{n})^\top$. Let $\beta_n(b)=(\beta_{0,n}+ \frac{b}{\sqrt{n}})/||\beta_{0,n}+ \frac{b}{\sqrt{n}}||$, with  $b= (b_1, \ldots, b_d)^\top\in \mathbb{Z}^d$ fixed. Recall that according to Assumption~\ref{a3}, $\check \beta_n$ is an estimator of $\beta_0$ such that $\sqrt{n}\check \beta_n\in \mathbb{Z}^d$  and  $\check \beta_n-\beta_0=O_\mathbb{P}(1/\sqrt{n})$; also  recall  $\widehat \beta=\check \beta_n/||\check \beta_n||$.
\begin{lemma}\label{Lemma 2}
Assume $$||\delta_n( \beta_n(b))||:=\sup_{u\in [0,1]}|\delta_n(u, \beta_n(b))|\stackrel{\mathbb{P}}{\to} 0,$$ for each $b \in \mathbb{Z}^d$. Then, as $n\to \infty$,
$$||\delta_n(\widehat \beta)||\stackrel{\mathbb{P}}{\to} 0.$$
\end{lemma}

\noindent{\bf Proof.}  Let $\varepsilon, M >0$. Then \begin{eqnarray*}&&\!\!\!\! \!\!\!\! \mathbb{P}(||\delta_n(\widehat \beta)||> \varepsilon)\\
&&= \mathbb{P}(||\delta_n(\widehat \beta)||> \varepsilon, \sqrt{n}||\check \beta_n-\beta_{0,n}||>M)
 +\mathbb{P}(||\delta_n(\widehat \beta)||> \varepsilon, \sqrt{n}||\check \beta_n-\beta_{0,n}||\leq M)\\
&&\leq \mathbb{P}( \sqrt{n}||\check \beta_n-\beta_{0,n}||>M)
 + \sum_{b\in \mathbb{Z}^d,\, || b||\leq M}\mathbb{P}(||\delta_n( \beta_n(b))||> \varepsilon). \end{eqnarray*}
Now for $M=M_\varepsilon$ large enough, the first probability is less than $\varepsilon/2$, for $n$ large enough. The second probability is, for large $n$, smaller than $\varepsilon/(2(2M+1)^d)$, which makes the second term less than  $\varepsilon/2$, too. $\hfill\Box$

Let $ \beta_n(b)$ be as above and
 define $A_k$ based on $ \beta_n(b)$. Recall  $ F_{j,k}(z) =\mathbb{P}(Z\leq z\, | \, X\in A_k, Y=j)$ and  $ F_k(z) =\mathbb{P}(Z\leq z\, | \, X\in A_k)$. Write  $p_{j,k}=\mathbb{P}(Y=j\, | \, X\in A_k)$,  $H_z(v)=\mathbb{P}(Z\leq z\,|\, X^\top \beta_0=v)$ and  $L_j(v)=\mathbb{P}(X^\top \beta \leq v\,|\, Y=j)$ with $\beta= \beta_n(b)$ or $\beta=\beta_0$.
\begin{lemma} \label{lem3} Let Assumption~\ref{a2} hold for $\beta=\beta_n(b)$ and assume $m(a_{k+1}-a_k)$ stays bounded, 
as $m\to \infty$. Assume that the conditional probability distribution of $X$ given $Y=j$ has, for $j=0,1$, a bounded density; that  $H_z$ is twice differentiable with uniformly, in $z$, bounded derivatives $H'_z$ and $H''_z$; and that, for all $\beta$ as above, $L_j$ is also twice differentiable with $L'_j$ bounded away from 0 (for values $v$ where   $0<L_j(v)<1$) 
and $L''_j$ bounded. Then, for $j=0,1$ and as $n \to \infty$,
$$\max_{k\in\{1, \ldots, m\}} \sup_{z\in \mathbb{R}}|F_{j,k}(z)-F_k(z)|=O(1/m^2).$$
\end{lemma} 
Note that in case $L'_j$ is not bounded away from zero at the endpoints of the support of $X^\top\beta$, one may restrict the analysis to observations away from these endpoints.\\

\noindent{\bf Proof.}
We have $F_{1,k}(z)-F_k(z)=p_{0,k}(F_{1,k}(z)-F_{0,k}(z))$ and $F_{0,k}(z)-F_k(z)=p_{1,k}(F_{0,k}(z)-F_{1,k}(z))$. Therefore, it suffices to show the result with $F_{j,k}(z)-F_k(z)$ replaced by $F_{1,k}(z)-F_{0,k}(z)$.

Now, writing $G_{j,k}(x)=\mathbb{P}(X\leq x\,|\, X\in A_k, Y=j)$, we have, for some $\theta$ between $a_k$ and $x^\top \beta_0$,
\begin{eqnarray*}&&F_{1,k}(z)-F_{0,k}(z)=\int_{A_k}\mathbb{P}(Z\leq z\,|\, X=x)d(G_{1,k}(x)-G_{0,k}(x))\\&&=\int_{A_k}H_z(x^\top \beta_0)d(G_{1,k}(x)-G_{0,k}(x))\\
&&=\int_{A_k}H_z(a_k)+H'_z(a_k)(x^\top \beta_0-a_k)+\frac{1}{2}H''_z(\theta)(x^\top \beta_0-a_k)^2d(G_{1,k}(x)-G_{0,k}(x)) \\
&&=H'_z(a_k)\int_{A_k}(x^\top \beta_0-a_k)d(G_{1,k}(x)-G_{0,k}(x))+\frac{1}{2}\int_{A_k}H''_z(\theta)(x^\top \beta_0-a_k)^2d(G_{1,k}(x)-G_{0,k}(x)).\end{eqnarray*}
Writing $x^\top\beta_0-a_k=x^\top\beta_n(b)-a_k+x^\top(\beta_0-  \beta_n(b))$, we obtain that the last term is $O((1/m +1/\sqrt{n})^2)=O(1/m^2)$.

Hence, it remains to show that $\int_{A_k}(x^\top\beta_0-a_k)d(G_{1,k}(x)-G_{0,k}(x))=O(1/m^2)$. We write this integral as $\int_{A_k}(x^\top \beta_n(b)-a_k)d(G_{1,k}(x)-G_{0,k}(x))+\int_{A_k}x^\top (\beta_0- \beta_n(b))d(G_{1,k}(x)-G_{0,k}(x))$. For the first integral, we assume w.l.o.g.\ $\beta_n(b)=(1,0, \ldots, 0)^\top$. Then
\begin{eqnarray*}&&\int_{A_k}(x^\top \beta_n(b)-a_k)d(G_{1,k}(x)-G_{0,k}(x))=\int_{A_k}(x_1-a_k)d(G_{1,k}(x)-G_{0,k}(x))\\&&\qquad=\int_{a_k}^{a_{k+1}}(x_1-a_k)d(L_{1,k}(x_1)-L_{0,k}(x_1)),\end{eqnarray*}
with
$L_{j,k}(x_1)=G_{j,k}(x_1, \infty, \dots, \infty)$. Now,
\begin{eqnarray*}L_{j,k}(x_1)&&=\frac{\mathbb{P}(X_1\leq x_1, X\in A_k\,|\, Y=j)}{\mathbb{P}(X\in A_k\,|\, Y=j)}=\frac{L_j(x_1)-L_j(a_k)}{L_j(a_{k+1})-L_j(a_k)}\\&&=\frac{(x_1-a_k)L'_j(a_k)+\frac{1}{2}(x_1-a_k)^2L''_j(\theta_1)}{(a_{k+1}-a_k)L'_j(a_k)+\frac{1}{2}(a_{k+1}-a_k)^2L''_j(\theta_2)}\\
&&=\frac{(x_1-a_k)/(a_{k+1}-a_k)+\frac{1}{2}(x_1-a_k)^2L''_j(\theta_1)/(L'_j(a_k)(a_{k+1}-a_k))}{1+\frac{1}{2}(a_{k+1}-a_k)L''_j(\theta_2)/L'_j(a_k)}\\
&&=\frac{(x_1-a_k)/(a_{k+1}-a_k)+O(1/m)}{1+O(1/m)}=(x_1-a_k)/(a_{k+1}-a_k)+O(1/m).\end{eqnarray*}
Hence, $L_{1,k}(x_1)-L_{0,k}(x_1)=O(1/m)$, uniformly in $x_1$ and $k$. Using this and integration by parts readily yields
$\int_{a_k}^{a_{k+1}}(x_1-a_k)d(L_{1,k}(x_1)-L_{0,k}(x_1))=O(1/m^2)$.

Hence, the proof is complete if we show  $\int_{A_k}x^\top (\beta_0- \beta_n(b))d(G_{1,k}(x)-G_{0,k}(x))=O(1/m^2)$.
For this we assume, again w.l.o.g., $\beta_0=(1,0.\ldots, 0)^\top$. Then $\beta_0- \beta_n(b) =\widetilde b/\sqrt{n}$ where  $\widetilde b=\widetilde b(n) =(\widetilde b_1, \ldots, \widetilde b_d)^\top$ stays  bounded.
Hence,
\begin{align*}
    &\int_{A_k}x^\top (\beta_0- \beta_n(b))d(G_{1,k}(x)-G_{0,k}(x)) \\
    &\qquad=\int_{A_k}\widetilde b_1 x_1/\sqrt{n}+\widetilde b_2 x_2/\sqrt{n}\dots+\widetilde b_d x_d/\sqrt{n} d(G_{1,k}(x)-G_{0,k}(x)).
\end{align*}
Now, since $m(\sup_{x\in A_k}x_1-\inf_{x\in A_k} x_1)$ stays bounded,  we have that the first term from the latter integral is $O(1/(\sqrt{n}m))=O(1/m^2)$. The other $d-1$ terms all can be handled in the same way. We, therefore, only consider the term with $x_2$.
Defining $p(A_k)$ as the projection of $A_k$ on the $x_1$-axis, it can be written as $$\frac{\widetilde b_2 }{\sqrt{n}}\int_{p(A_k)\times\mathbb{R}^{d-1}} x_2 d(G_{1,k}(x)-G_{0,k}(x)).$$
Define $I_k=\{x_1:  x\in S, x_1\in p(A_k) \mbox{ implies } x \in A_k\}.$
Then, for $j=0,1$,
\begin{eqnarray*}&&\left|\frac{\widetilde b_2 }{\sqrt{n}}\int_{(p(A_k)\setminus I_k)\times\mathbb{R}^{d-1}} x_2 dG_{j,k}(x)\right|=O\left(\frac{m}{\sqrt{n}}\right)\int_{(p(A_k)\setminus I_k)\times\mathbb{R}^{d-1}} |x_2| d\mathbb{P}(X\leq x\,|\, Y=j)\\&&\qquad=O\left(\frac{m}{\sqrt{n}}\right)O\left(\frac{1}{\sqrt{n}}\right)=O\left(\frac{m}{n}\right)=O\left(\frac{1}{m^2}\right).\end{eqnarray*}
Now it remains to consider $$\frac{\widetilde b_2 }{\sqrt{n}}\int_{I_k\times\mathbb{R}^{d-1}} x_2 d(G_{1,k}(x)-G_{0,k}(x)).$$ Define $\widetilde G_{j,k}(x)=\mathbb{P}(X\leq x\,|\, X_1\in I_k, Y=j). $ Observe  that, on $I_k\times \mathbb{R}^{d-1}$, the density of  $ G_{j,k}$ is a multiple of that of  $\widetilde G_{j,k}$ and 1 minus  this multiplication factor is $O(m/\sqrt{n})$.
Then $$\left|\frac{\widetilde b_2 }{\sqrt{n}}\int_{I_k\times\mathbb{R}^{d-1}} x_2 d(G_{j,k}(x)-\widetilde G_{j,k}(x))\right|=O\left(\frac{1}{\sqrt{n}}\right)O\left(\frac{m}{\sqrt{n}}\right)=O\left(\frac{m}{n}\right)=O\left(\frac{1}{m^2}\right).$$
Hence, it finally remains to show  $$\frac{\widetilde b_2 }{\sqrt{n}}\int_{I_k\times\mathbb{R}^{d-1}} x_2 d(\widetilde G_{1,k}(x)-\widetilde G_{0,k}(x))=O\left(\frac{1}{m^2}\right).$$
This integral expression  is equal to  \begin{eqnarray*}&&\frac{\widetilde b_2 }{\sqrt{n}}\int_{I_k\times\mathbb{R}^{d-1}} x_2 d G_{*,x_1}(x_2, \ldots, x_d)d(\check G_{1,k}(x_1)-\check G_{0,k}(x_1))\\
&&\qquad=\frac{\widetilde b_2 }{\sqrt{n}}\int_{I_k}E (X_2\,|\, X_1=x_1) d(\check G_{1,k}(x_1)-\check G_{0,k}(x_1)),\end{eqnarray*}
with $\check G_{j,k}(x_1)=\mathbb{P}(X_1\leq x_1\,|\, X_1\in I_k, Y=j)$ and $ G_{*,x_1}(x_2, \ldots, x_d)=\mathbb{P}(X_2\leq x_2, \dots, X_d\leq x_d\,|\, X_1=x_1, Y=j)$; the * indicates that the distribution function does not depend on $j$, by assumption (\ref{eq:SingleIndexY}). We have again, as for $L_{1,k}-L_{0,k}$ above, $\check G_{1,k}(x_1)-\check G_{0,k}(x_1)=O(1/m)$, uniformly in $x_1$ and $k$. Integration by parts now yields that the last integral expression is $O(1/\sqrt{n}) O(1/m)=O(1/m^2)$.$\hfill\Box$

\vspace{0.3 cm}

We now write $\gamma_n(\cdot , \beta)$  when the process $\gamma_n$ in (\ref{gam}), depends on a directional vector $\beta$, not necessarily $\widehat \beta$. In particular, we consider  $\beta=\beta_n(b)$, as in Lemma \ref{Lemma 2}.  Write $q_j=\mathbb{P}(Y=j)$, $q_{j,k}=\mathbb{P}(X\in A_k, Y=j)$ and \begin{equation}\label{cn} c_n=\sqrt{1-\frac{1}{q_0q_1}\sum_{k=1}^{m}\frac{(q_{1}q_{0,k}-q_{0}q_{1,k})^2}{q_{0,k}+q_{1,k}}}\,.\end{equation}
Define, for $u\in [0,1]$, 
\begin{eqnarray}&&\widetilde B_n(u, \beta_n(b)) = \frac{\sqrt{q_0q_1}}{c_n}\sum_{k=1}^{m}\left\{\left[\left(\frac{q_{1,k}}{q_1}-\frac{q_{0,k}}{q_0}\right)
\frac{\sqrt{ q_{0,k} }}{q_{0,k}+q_{1,k} }\right.
+\frac{\sqrt{ q_{0,k} }}{q_0 }\right]B_{0,k,n}(u) \nonumber\\
&&\left. \qquad \qquad \qquad \qquad \qquad 
+
\left[\left(\frac{q_{1,k}}{q_1}-\frac{q_{0,k}}{q_0}\right)
\frac{\sqrt{ q_{1,k} }}{q_{0,k}+q_{1,k} }
-\frac{\sqrt{ q_{1,k} }}{q_1 }\right]B_{1,k,n}(u)\right\} , \label{consti}\end{eqnarray}
with the $B_{j,k,n}$ from Proposition~\ref{prop1}. Direct covariance calculations yield that for every $b$, $\widetilde B_n(\cdot, \beta_n(b))$ is a standard Brownian bridge.
\begin{proposition} \label{bee}
We have, under the assumptions of Theorem~\ref{t1} and as $n\to \infty$,
\begin{equation}\label{prtwo}\sup_{u\in [0,1]}| \gamma_n(u, \beta_n(b))-\widetilde B_n(u, \beta_n(b))|\stackrel{\mathbb{P}}{\to} 0.\end{equation}
\end{proposition}

\noindent{\bf Proof.} Let
 $\widehat F_k^{-1} $ be the (left-continuous) generalized inverse of $\widehat F_k$. Also set $w_{j,k}=n_{j,k}/n_j$.
Then $$ \widehat \Gamma_j(u)=\sum_{k=1}^m w _{j,k}\widehat F_{j,k}(\widehat F_k^{-1}(u)), \quad u\in [0,1], \, j=0,1. $$
Define 
\begin{eqnarray*}&&\breve B_n(u) =
\sqrt{\frac{n_0n_1}{n}}\left( \sum_{k=1}^m w _{0,k}F_{0,k}(\widehat F_k^{-1}(u)) -\sum_{k=1}^m w _{1,k} F_{1,k}(\widehat F_k^{-1}(u))\right)\\
&&\qquad +
\sqrt{\frac{n_0n_1}{n}}\left( \sum_{k=1}^m w _{0,k}B_{0,k,n}(F_{0,k}(\widehat F_k^{-1}(u))/\sqrt{n_{0,k}} -\sum_{k=1}^m w _{1,k} B_{1,k,n}(F_{1,k}(\widehat F_k^{-1}(u))/\sqrt{n_{1,k}} \right).\end{eqnarray*} 
Now Proposition \ref{prop1} yields
\begin{equation}\label{firs} \sup_{u\in [0,1]} \left| \sqrt{\frac{n_0n_1}{n}}\,\left(
	\widehat{\Gamma}_0(u)-\widehat{\Gamma}_1(u)\right)-\breve B_n(u)\right|=O_\mathbb{P}\left( \frac{m}{\sqrt{n}} \log n\right).\end{equation}
Observe that $(m/\sqrt{n}) \log n= ((m^3 \log^3 n)/n^{3/2})^{1/3} \to 0$ as $n\to \infty$.
We use Lemma \ref{lem3} to replace $F_{0,k}$ and $F_{1,k}$ by $F_k$ in the definition of $\breve B_n$.
Define
\begin{eqnarray}\nonumber
\check B_n(u) &=&
\sqrt{\frac{n_0n_1}{n}}\left( \sum_{k=1}^m (w _{0,k} -w _{1,k}) F_{k}(\widehat F_k^{-1}(u))\right)\label{check}\\ \nonumber
 & & +
\sqrt{\frac{n_0n_1}{n}}\left( \sum_{k=1}^m w _{0,k}B_{0,k,n}(F_{k}(\widehat F_k^{-1}(u))/\sqrt{n_{0,k}} -\sum_{k=1}^m w _{1,k} B_{1,k,n}(F_{k}(\widehat F_k^{-1}(u))/\sqrt{n_{1,k}} \right).\end{eqnarray} 
Lemma~\ref{lem3} yields 
\begin{equation}\label{seco}\sup_{u\in [0,1]} \left| \breve B_n(u)-\check B_n(u)\right|=
O_\mathbb{P}\left( \frac{\sqrt{n}}{m^2} +\sqrt {\frac{\log m}{m}}\right)=o_\mathbb{P}(1). \end{equation} Here we have used that, if the inputs are perturbed by order $1/m^2$,  the oscillation of the $2m$ Brownian bridges is bounded uniformly by order $\sqrt{\log m}/m$, in probability,
which follows from standard probability inequalities for Brownian bridges.

Write 
$$F_{k}(\widehat F_k^{-1}(u))=u+\frac{\sqrt{n_{0,k}+n_{1,k}}(F_{k}(\widehat F_k^{-1}(u))-u)}{\sqrt{n_{0,k}+n_{1,k}}}=: u+\frac{\delta_k(u)}{\sqrt{n_{0,k}+n_{1,k}}}$$ and note that the $\delta_k$ are uniform quantile processes.
Similarly, write
$$\widehat F_{k} (F_k^{-1}(u))=u+\frac{\sqrt{n_{0,k}+n_{1,k}}(\widehat F_{k}( F_k^{-1}(u))-u)}{\sqrt{n_{0,k}+n_{1,k}}}=: u+\frac{\alpha_k(u)}{\sqrt{n_{0,k}+n_{1,k}}}$$ and note that the $\alpha_k$ are uniform empirical processes.
Now classical empirical process theory, see \cite{shorack2009empirical},
yields
$$\max_{k\in\{1, \ldots, m\}}\sup_{u\in [0,1]} \left| \alpha_k(u)+\delta_k(u)\right|=
O_\mathbb{P}\left((\log n)^{3/4} m^{1/4}/n^{1/4} \right);$$ this is a so-called Bahadur-Kiefer result. It yields
$$\nonumber\sup_{u\in [0,1]}\left|\sqrt{\frac{n_0n_1}{n}}\left( \sum_{k=1}^m (w _{0,k} -w _{1,k}) F_{k}(\widehat F_k^{-1}(u))\right)- \sqrt{\frac{n_0n_1}{n}}\left( \sum_{k=1}^m (w _{1,k} -w _{0,k})\widehat F_{k}( F_k^{-1}(u))\right)\right|
$$
\begin{eqnarray}&&= O_\mathbb{P}\left(n^{1/2}(\log n)^{3/4} \frac{m^{1/4}}{n^{1/4}}  \frac{m^{1/2}}{n^{1/2}}\right)
=O_\mathbb{P}\left(\left((\log n)^3\frac{m^3}{n} \right)^{1/4} \right)
=o_\mathbb{P}(1).\label{ninet}\end{eqnarray}
Applying Proposition \ref{prop1} and twice Lemma \ref{lem3}, we obtain
\begin{eqnarray}\label{twen}&&\nonumber\sup_{u\in [0,1]}\left| \sqrt{\frac{n_0n_1}{n}}\left( \sum_{k=1}^m (w _{1,k} -w _{0,k})\widehat F_{k}( F_k^{-1}(u))\right)\right.\\&&\qquad\qquad\left.- \sqrt{\frac{n_0n_1}{n}}\left( \sum_{k=1}^m (w _{1,k} -w _{0,k})\left[
\frac{\sqrt{n_{0,k}}B_{0,k,n} (u)} {n_{0,k}+n_{1,k}}
+\frac{\sqrt{n_{1,k}}B_{1,k,n} (u)} {n_{0,k}+n_{1,k}}\right]\right)\right|\nonumber
\\&&
=O_\mathbb{P}\left(\left(\frac{m^2}{n}\log^2 n \right)^{1/2} +\frac{\sqrt{n}}{m}+\sqrt{\frac{\log m}{m}}\right)
=o_\mathbb{P}(1).\end{eqnarray}
Next we approximate the second term in (\ref{check}), by replacing $F_{k}(\widehat F_k^{-1}(u)) $ by $u$.
Write
\begin{eqnarray*}\label{twen1}
&&\overline B_n(u)=\sqrt{\frac{n_0n_1}{n}}\left( \sum_{k=1}^m (w _{1,k} -w _{0,k})\left[
\frac{\sqrt{n_{0,k}}B_{0,k,n} (u)} {n_{0,k}+n_{1,k}}
+\frac{\sqrt{n_{1,k}}B_{1,k,n} (u)} {n_{0,k}+n_{1,k}}\right]\right)\\&&
\qquad\qquad\qquad +
\sqrt{\frac{n_0n_1}{n}}\left( \sum_{k=1}^m w _{0,k}B_{0,k,n}(u)/\sqrt{n_{0,k}} -\sum_{k=1}^m w _{1,k} B_{1,k,n}(u)/\sqrt{n_{1,k}} \right).
\end{eqnarray*}
We have $\sup_{u\in [0,1]}|F_{k}(\widehat F_k^{-1}(u))-u|=O_\mathbb{P}(\sqrt{(m \log n)/n}) $. 
Using this  and (\ref{ninet}) and (\ref{twen}), we obtain
\begin{eqnarray}\label{twen2}&&\nonumber
\sup_{u\in [0,1]}\left|\check B_n(u)-\overline B_n(u)\right|=O_\mathbb{P}\left(n^{1/2}
\left(\frac{m}{n}\right)^{1/2} \left(\frac{m}{n} \log n\right)^{1/4} (\log n)^{1/2}\right)+o_\mathbb{P}(1)\\&&\qquad\qquad=O_\mathbb{P}\left(\left(\frac{m^3}{n} \log^3 n\right)^{1/4}\right)+o_\mathbb{P}(1)=o_\mathbb{P}(1).
\end{eqnarray}

Multiple applications of the WLLN now yield
\begin{equation}\label{twen3}\sup_{u\in [0,1]}\left|\overline B_n(u)-c_n\widetilde B_n(u, \beta_n(b)) \right|=o_\mathbb{P}(1).\end{equation}
Note that $\liminf_{n\to\infty}  c_n>0$. Now combining (\ref{firs}), (\ref{seco}), (\ref{twen2}), (\ref{twen3}), and again multiple applications of the WLLN, yields~(\ref{prtwo}).
$\hfill\Box$

In order to present our final proposition, we need to explicitly define 
the Brownian bridges $B_{0,k,n}$ and $B_{1,k,n}$ in Proposition \ref{prop1}, which are the building blocks of $\widetilde B_n(u, \beta_n(b))$, see (\ref{consti}).   Note that these Brownian bridges are  minor modifications of the approximating Brownian bridges in the Proposition in \cite{beirlant1996maximal}.

Recall that  $S$ denotes the support of $X$. Define the transformation $T$ from $S\times 
\{0,1\} \times \mathbb{R} $ to $[0,1]$ as follows
\begin{eqnarray*} 
&& T(x,0, z):=\sum_{k=1}^m I_{A_k}(x)\left[\sum_{\ell=1}^{k-1}q_{0,\ell}+q_{0,k}F_{0,k}(z)\right],\\
&& T(x,1, z):=q_0+\sum_{k=1}^m I_{A_k}(x)\left[\sum_{\ell=1}^{k-1}q_{1,\ell}+q_{1,k}F_{1,k}(z)\right].
\end{eqnarray*}
We also define, for $j=0,1$, the set 
$$T_{j,k}(z):= \{T(x, j, v): x\in A_k, v\leq z\}, \quad z \in \mathbb{R} \cup \{\infty\}.$$ Consider a given sequence
$W_n$, $n\in \mathbb{N}$,  of standard Wiener processes on $[0,1]$ and write for an interval $A$, being $[a,c]$ or $(a,c]$ or $(a,c)$, 
$W_n(A)=W_n(c)-W_n(a)$.
Define for $z \in \mathbb{R} $,
\begin{eqnarray}\label{be} &&\nonumber B_{j,k,n}(F_{j,k}(z)):=\frac{1}{\sqrt{q_{j,k}}}\{
W_n(T_{j,k}(z) )-F_{j,k}(z)
W_n(T_{j,k}(\infty) )\}.\end{eqnarray}
Note that here the  dependence on the vector $b$ is still suppressed in the  notation.

\begin{proposition} \label{beenull}
We have, under the assumptions of Theorem~\ref{t1} and as $n\to \infty$,
\begin{equation}\label{prt}\sup_{u\in [0,1]}| \widetilde B_n(u, \beta_n(b))- \widetilde B_n(u, \beta_n(\mathbf{0}))|\stackrel{\mathbb{P}}{\to} 0, \quad \mbox{with } \mathbf{0}=(0, \ldots, 0)^\top. \end{equation}
\end{proposition}

\noindent{\bf Proof.} We need to show that the random function
$\widetilde B_n(\cdot\, , \beta_n(b))- \widetilde B_n(\cdot\, , \beta_n(\mathbf{0}))$ converges weakly on $C[0,1]$ to the 0-function.
Since the random function is the difference of two standard Brownian bridges, the asymptotic tightness is well-known. Hence, it remains to show the convergence of the finite-dimensional distributions. As the limit is the 0-function, for this it is sufficient to show the convergence to 0 at a single, arbitrary point $u$. We will prove this by showing, as $n\to \infty$,
$$\var( \widetilde B_n(u, \beta_n(b))- \widetilde B_n(u, \beta_n(\mathbf{0})))\to 0.$$

For a vector $\beta$ with $\|\beta\| = 1$, write
\begin{align*}
    c_{0,k}(\beta) &\coloneq \frac{q_{1,k}(\beta)/q_1-q_{0,k}(\beta)/q_0}{q_{0,k}(\beta) + q_{1,k}(\beta)} + \frac{1}{q_0}, \\
    c_{1,k}(\beta) &\coloneq \frac{q_{1,k}(\beta)/q_1-q_{0,k}(\beta)/q_0}{q_{0,k}(\beta) + q_{1,k}(\beta)} - \frac{1}{q_1},
\end{align*}
so that we can write
\begin{equation*}\label{d}
    \widetilde{B}_n(u, \beta)= \frac{\sqrt{q_0q_1}}{c_n}\sum_{k=1}^m \sum_{j\in \{0,1\}} c_{j,k}(\beta) D_{j,k,n}(u, \beta),
\end{equation*}
where $D_{j,k,n}(u, \beta)$, $j\in \{0,1\}$, $k=1,\ldots, m$, are independent Brownian bridges, each with variance $q_{j,k}(\beta)u(1-u)$, respectively,  and $c_n=c_n(\beta)$, see (\ref{cn}). We write
\[
   \widetilde B_n(u, \beta_n(b))- \widetilde B_n(u, \beta_n(\mathbf{0}))= \xi_{1,n}(u)+\xi_{2,n}(u)+\xi_{3,n}(u),
\]
where
\begin{align*}
    \xi_{1,n}(u) &\coloneq \sqrt{q_0q_1}\left(\frac{1}{c_n(\beta_n(b))}-\frac{1}{c_n(\beta_n(\mathbf{0}))}\right)\sum_{k=1}^m \sum_{j\in \{0,1\}} c_{j,k}(\beta_n(b)) D_{j,k,n}(u, \beta_n(b)), \\
    \xi_{2,n}(u) &\coloneq \frac{\sqrt{q_0q_1}}{c_n(\beta_n(\mathbf{0}))}\sum_{k=1}^m\sum_{j\in \{0,1\}} (c_{j,k}(\beta_n(b))-c_{j,k}(\beta_n(\mathbf{0}))) D_{j,k,n}(u, \beta_n(b)),  \\
    \xi_{3,n}(u) &\coloneq \frac{\sqrt{q_0q_1}}{c_n(\beta_n(\mathbf{0}))}\sum_{k=1}^m \sum_{j\in \{0,1\}} c_{j,k}(\beta_n(\mathbf{0}))(D_{j,k,n}(u, \beta_n(b))-D_{j,k,n}(u, \beta_n(\mathbf{0}))).
\end{align*}
It is readily shown that
uniformly for $j\in \{0,1\}$ and $1\le k\le m$,
\begin{equation} \label{bond}
    |q_{j,k}(\beta_n(b)) - q_{j,k}(\beta_n(\mathbf{0}))|=O(  \|\beta_n(b)-\beta_n(\mathbf{0})\|).
\end{equation}
We use this for handling  $\xi_{1,n}$ and $ \xi_{2,n}$. We begin with $ \xi_{2,n}$.



Let
\begin{align*}
    c_0(x,y)&\coloneq \frac{y/q_1-x/q_0}{x+y} + \frac{1}{q_0}, \\
    c_1(x,y)&\coloneq \frac{y/q_1-x/q_0}{x+y} - \frac{1}{q_1},
\end{align*}
and note that for $x,y>0$,
\[
    |\partial_x c_j(x,y)|+|\partial_y c_j(x,y)|=\frac{q_0^{-1}+q_1^{-1}}{x+y}.
\]
Using the mean-value theorem, we find that, with $a>0$ a lower bound on the $q_{j,k}(\,\cdot\,)$,
\begin{eqnarray*}&&\hspace{-0.7cm} 
    |c_{j,k}(\beta_n(b)) - c_{j,k}(\beta_n(\mathbf{0}))| = |c_j(q_{0,k}(\beta_n(b)), q_{1,k}(\beta_n(b))) - c_j(q_{0,k}(\beta_n(\mathbf{0})), q_{1,k}(\beta_n(\mathbf{0})))| \\
    &&\hspace{-0.7cm}\leq 2\sup_{(x,y)\in [a,1]^2}\left\{|\partial_x c_j(x,y)|+|\partial_y c_j(x,y)|\right\} 
     \left\{|q_{0,k}(\beta_n(b))-q_{0,k}(\beta_n(\mathbf{0}))|+|q_{1,k}(\beta_n(b))-q_{1,k}(\beta_n(\mathbf{0}))|\right\}.
\end{eqnarray*}
Since  $m\cdot   q_{j,k}(\,\cdot\,)$, is bounded away from $0$ uniformly  for large  $m$, we obtain from (\ref{bond})
\begin{align*}
    |c_{j,k}(\beta_n(b)) - c_{j,k}(\beta_n(\mathbf{0}))| = O(m\cdot \|\beta_n(b)-\beta_n(\mathbf{0})\|).
\end{align*}
Now, 
\begin{eqnarray*}&&
    \var(\xi_{2,n}(u)) 
    =\frac{q_0q_1}{c_n^2(\beta_n(\mathbf{0}))}\sum_{k=1}^m \sum_{j\in \{0,1\}} (c_{j,k}(\beta_n(b))-c_{j,k}(\beta_n(\mathbf{0})))^2 \var(D_{j,k,n}(u,\beta_n(b))) \\
    &&= O(m^3\cdot \|\beta_n(b)-\beta_n(\mathbf{0})\|^2)=O(m^3/n)=o(1),
\end{eqnarray*}
because $c_n(\beta_n(\mathbf{0}))$ is bounded away from $0$ for large $n$ and $\var(D_{j,k,n}(u))\le u(1-u)$.

Next we consider the variance of $\xi_{1,n}(u)$.
Letting
\[
    g(x,y)\coloneq \frac{(q_1 x - q_0 y)^2}{x+y}\, ,
\]
we have
\begin{align*}
    |c_n^2(\beta_n(b))-c_n^2(\beta_n(\mathbf{0}))|=\frac{1}{q_0q_1}\left|\sum_{k=1}^m g(q_{0,k}(\beta_n(b)),q_{1,k}(\beta_n(b)))-g(q_{0,k}(\beta_n(\mathbf{0})),q_{1,k}(\beta_n(\mathbf{0})))\right|.
\end{align*}
Inspecting the partial derivatives of $g$ on $(0,1]^2$  yields
\begin{align*}
    |\partial_x g(x,y)|&\le \frac{2q_0|q_1x-q_0y|}{x+y}+\frac{|q_1x-q_0y|^2}{(x+y)^2} \le 3\max\{q_0^2,q_1^2\},\\
    |\partial_y g(x,y)|&\le \frac{2q_1|q_1x-q_0y|}{x+y}+\frac{|q_1x-q_0y|^2}{(x+y)^2}\le 3\max\{q_0^2,q_1^2\}.
\end{align*}
Thus, since $c_n(\,\cdot\,)$ is bounded away from $0$ for $n$ large enough, we obtain (similarly to the previous case)
\[
    |c_n(\beta_n(b))-c_n(\beta_n(\mathbf{0}))|=\frac{|c_n^2(\beta_n(b))-c_n^2(\beta_n(\mathbf{0}))|}{c_n(\beta_n(b))+c_n(\beta_n(\mathbf{0}))}=O(m\cdot \|\beta_n(b)-\beta_n(\mathbf{0})\|).
\]
Since $|c_{j,k}(\,\cdot\,)|\le 2\max\{q_0^{-1},q_1^{-1}\}$, we find
\begin{align*}
    &\var(\xi_{1,n}(u)) 
    =q_0q_1\left(\frac{c_n(\beta_n(\mathbf{0}))-c_n(\beta_n(b))}{c_n(\beta_n(b))c_n(\beta_n(\mathbf{0}))}\right)^2\sum_{k=1}^m \sum_{j\in \{0,1\}} c^2_{j,k}(\beta_n(b)) \var(D_{j,k,n}(u, \beta_n(b))) \\
    &\quad= O(m^3\cdot \|\beta_n(b)-\beta_n(\mathbf{0})\|^2)=O(m^3/n)=o(1).
\end{align*}

Finally, consider the variance of $\xi_{3,n}(u)$.
Note that for showing that this variance tends to 0, $\frac{\sqrt{q_0q_1}}{c_n(\beta_n(\mathbf{0}))}$ can be ignored. Moreover, it suffices to study $j=0,1$ separately; we consider the case $j=0$ only; $j=1$ can be dealt with similarly.
We have, see~(\ref{be}),
\begin{eqnarray}
	&&\qquad \sum_{k=1}^m c_{0,k}(\beta_n(\mathbf{0}))\left[D_{0,k,n}(u, \beta_n(b))- D_{0,k,n}(u, \beta_n(\mathbf{0}))\right]\label{eq:proofPart2}\\
	\nonumber
	&&\qquad\quad  =\sum_{k=1}^m c_{0,k}(\beta_n(\mathbf{0})) \left[W_n\left(\sum_{l=1}^{k-1}q_{0l}(\beta_n(b))+q_{0k}(\beta_n(b))u\right)- W_n\left(\sum_{l=1}^{k-1}q_{0l}(\beta_n(b))\right)\right.\\
	\nonumber
	&&\qquad\qquad-\left.\left\{W_n\left(\sum_{l=1}^{k-1}q_{0l}(\beta_n(\mathbf{0}))+q_{0k}(\beta_n(\mathbf{0}))u\right)- W_n\left(\sum_{l=1}^{k-1}q_{0l}(\beta_n(\mathbf{0}))\right)\right\}\right.\\
	\nonumber
	&&\qquad\qquad-u\left(W_n\left(\sum_{l=1}^{k}q_{0l}(\beta_n(b))\right)- W_n\left(\sum_{l=1}^{k-1}q_{0l}(\beta_n(b))\right)\right.\\
	\nonumber
	&&\qquad\qquad-\left.\left.\left\{W_n\left(\sum_{l=1}^{k}q_{0l}(\beta_n(\mathbf{0}))\right)- W_n\left(\sum_{l=1}^{k-1}q_{0l}(\beta_n(\mathbf{0}))\right)\right\}\right)\right].
\end{eqnarray}
As $\max_{k=1,\ldots,m}\left|q_{0k}(\beta_n(b))-q_{0k}(\beta_n(\mathbf{0}))\right| = o\left(1/\sqrt{n}\right)$, we know that, for  $n$ large enough, the intervals $\left[\sum_{l=1}^{k'-1}q_{0l}(\beta_n(b)),\sum_{l=1}^{k'}q_{0l}(\beta_n(b))\right]$ and $\left[\sum_{l=1}^{k-1}q_{0l}(\beta_n(\mathbf{0})),\sum_{l=1}^{k}q_{0l}(\beta_n(\mathbf{0}))\right]$ are disjoint in case $k\neq k'$. As a result, when calculating the variance of~(\ref{eq:proofPart2}), cross-terms with $k'\neq k$ may be ignored. Also observe that the ``bridging operation'' on the last two lines of (\ref{eq:proofPart2}) reduces the variance relative to the variance of the Wiener process, which is maximal at $u=1$. As a result, the variance of the left-hand side of (\ref{eq:proofPart2}) can be bounded by
\begin{align*}
	&\sum_{k=1}^m c_{0,k}(\beta_n(\mathbf{0}))^2\var\left\{W_n\left(\sum_{l=1}^{k}q_{0l}(\beta_n(b))\right)- W_n\left(\sum_{l=1}^{k-1}q_{0l}(\beta_n(b))\right)\right.\\
	&~~~~~~~~~~~~~~~~~~~~~~~~~~~-\left.\left[W_n\left(\sum_{l=1}^{k}q_{0l}(\beta_n(\mathbf{0}))\right)- W_n\left(\sum_{l=1}^{k-1}q_{0l}(\beta_n(\mathbf{0}))\right)\right]\right\}\\
    &=\sum_{k=1}^m c_{0,k}(\beta_n(\mathbf{0}))^2\left[q_{0k}(\beta_n(b))+q_{0k}(\beta_n(\mathbf{0}))\right.\\
    &~~~~~~~~~~~~~~-2\left. \max\left\{0,\sum_{l=1}^{k}q_{0l}(\beta_n(b))\wedge\sum_{l=1}^{k}q_{0l}(\beta_n(\mathbf{0})) -\sum_{l=1}^{k-1}q_{0l}(\beta_n(b))\vee\sum_{l=1}^{k-1}q_{0l}(\beta_n(\mathbf{0}))\right\}\right]\\
	&=O\left(\frac{1}{\sqrt{n}}\right)\sum_{k=1}^m c_{0,k}(\beta_n(\mathbf{0}))^2
    =O\left(\frac{m}{\sqrt{n}}\right)=o(1),
\end{align*}
since $c_{0,k}(\beta_n(\mathbf{0}))^2\le (1/q_1+1/q_0)^2$. 

Since the variance of each  term $\xi_{1,n}(u), \xi_{2,n}(u)$ and $\xi_{3,n}(u)$ tends to 0, we have that
 $\var(\widetilde B_n(u, \beta_n(b))- \widetilde B_n(u, \beta_n(\mathbf{0}))) \to 0$, as $n\to \infty$.
$\hfill\Box$

\vspace{0.3 cm}

\noindent{\bf Proof of Theorem~\ref{t1}.} Combining Propositions~\ref{bee} and~\ref{beenull} and Lemma~\ref{Lemma 2}, with $$\delta_n(\beta_n(b)):=\gamma_n(\cdot\, ,\beta_n(b))-\widetilde B_n(\cdot\, , \beta_n(\mathbf{0})),$$ yields 
$\gamma_n =\gamma_n(\cdot \, , \widehat\beta)\rightsquigarrow B$. The convergence in distribution of the test statistic follows from the continuous mapping theorem.
$\hfill\Box$

\bibliographystyle{imsart-nameyear}
\bibliography{refs}

@article{beirlant1996maximal,
  title={Maximal Type Test Statistics based on Conditional Processes},
  author={Beirlant, Jan and Einmahl, J.H.J.},
  journal={Journal of Statistical Planning and Inference},
  volume={53},
  number={1},
  pages={1--19},
  year={1996},
  publisher={Elsevier}
}

@ARTICLE{berrett_2020,
     AUTHOR = {Berrett, Thomas B. and Wang, Yi and Barber, Rina Foygel and Samworth, Richard J.},
    JOURNAL = {Journal of the Royal Statistical Society: Series B (Statistical Methodology)},
     NUMBER = {1},
      PAGES = {175--197},
  PUBLISHER = {Wiley},
      TITLE = {The Conditional Permutation Test for Independence While Controlling for Confounders},
     VOLUME = {82},
       YEAR = {2020}
}

@ARTICLE{cai_li_zhang_2022,
     AUTHOR = {Cai, Zhanrui and Li, Runze and Zhang, Yaowu},
    JOURNAL = {Journal of Machine Learning Research},
     NUMBER = {85},
      PAGES = {1--41},
      TITLE = {A Distribution Free Conditional Independence Test with Applications to Causal Discovery},
     VOLUME = {23},
       YEAR = {2022}
}

@ARTICLE{li_fan_2019,
     AUTHOR = {Li, Cheng and Fan, Xiaodan},
    JOURNAL = {Wiley Interdisciplinary Reviews: Computational Statistics},
     NUMBER = {3},
      PAGES = {e1489},
  PUBLISHER = {Wiley},
      TITLE = {On Nonparametric Conditional Independence Tests for Continuous Variables},
     VOLUME = {12},
       YEAR = {2019}
}

@ARTICLE{powell_1989,
     AUTHOR = {Powell, James L. and Stock, James H. and Stoker, Thomas M.},
    JOURNAL = {Econometrica},
     NUMBER = {6},
      PAGES = {1403--1430},
  PUBLISHER = {JSTOR},
      TITLE = {Semiparametric Estimation of Index Coefficients},
     VOLUME = {57},
       YEAR = {1989}
}

@ARTICLE{shah_peters_2020,
     AUTHOR = {Shah, Rajen D. and Peters, Jonas},
    JOURNAL = {The Annals of Statistics},
     NUMBER = {3},
      PAGES = {1514--1538},
  PUBLISHER = {Institute of Mathematical Statistics},
      TITLE = {The Hardness of Conditional Independence Testing and the Generalised Covariance Measure},
     VOLUME = {48},
       YEAR = {2020}
}

@book{shorack2009empirical,
  title={Empirical Processes with Applications to Statistics},
  author={Shorack, Galen R and Wellner, Jon A},
  year={2009},
  publisher={SIAM}
}

@ARTICLE{song_2009,
     AUTHOR = {Song, Kyungchul},
    JOURNAL = {The Annals of Statistics},
     NUMBER = {6B},
      PAGES = {4011--4045},
  PUBLISHER = {Institute of Mathematical Statistics},
      TITLE = {Testing Conditional Independence Via {R}osenblatt Transforms},
     VOLUME = {37},
       YEAR = {2009}
}

@ARTICLE{su_white_2007,
     AUTHOR = {Su, Liangjun and White, Halbert},
    JOURNAL = {Journal of Econometrics},
     NUMBER = {2},
      PAGES = {807--834},
  PUBLISHER = {Elsevier},
      TITLE = {A Consistent Characteristic Function-Based Test for Conditional Independence},
     VOLUME = {141},
       YEAR = {2007}
}

@ARTICLE{su_white_2008,
     AUTHOR = {Su, Liangjun and White, Halbert},
    JOURNAL = {Econometric Theory},
     NUMBER = {4},
      PAGES = {829--864},
  PUBLISHER = {Cambridge University Press},
      TITLE = {A Nonparametric {H}ellinger Metric Test for Conditional Independence},
     VOLUME = {24},
       YEAR = {2008}
}

\end{document}